# Voltage-Gate Assisted Spin-Orbit Torque Magnetic Random Access Memory for High-Density and Low-Power Embedded Application

Y. C. Wu,[1,*] K. Garello,[1,2] W. Kim,[1] M. Gupta,[1] M. Perumkunnil,[1] V. Kateel,[1,3] S. Couet,[1] R. Carpenter,[1] S. Rao,[1] S. Van Beek,[1] K. K. Vudya Sethu,[1,3] F. Yasin,[1] D. Crotti,[1] and G. S. Kar[1]

[1]*imec, 3001 Leuven, Belgium*
[2]*Spintec, Grenoble, France*
[3]*Department of Electrical Engineering (ESAT), KU Leuven, 3001 Leuven, Belgium*
[*]e-mail: jackson_wij@yahoo.com.tw

Voltage-gate assisted spin-orbit torque (VGSOT) writing scheme combines the advantages from voltage control of magnetic anisotropy (VCMA) and spin-orbit torque (SOT) effects, enabling multiple benefits for magnetic random access memory (MRAM) applications. In this work, we give a complete description of VGSOT writing properties on perpendicular magnetic tunnel junction (pMTJ) devices, and we propose a detailed methodology for its electrical characterization. The impact of gate assistance on the SOT switching characteristics are investigated using electrical pulses down to 400ps. The VCMA coefficient ($\xi$) extracted from current switching scheme is found to be the same as that from the magnetic field switch method, which is in the order of 15fJ/Vm for the 80nm to 150nm devices. Moreover, as expected from the pure electronic VCMA effect, $\xi$ is revealed to be independent of the writing speed and gate length. We observe that SOT switching current characteristics are modified linearly with gate voltage ($V_g$), similar as for the magnetic properties. We interpret this linear behavior as the direct modification of perpendicular magnetic anisotropy (PMA) and nucleation energy induced by VCMA. At $V_g$ = 1V, the SOT write current is decreased by 25%, corresponding to a 45% reduction in total energy down to 30fJ/bit at 400ps speed for the 80nm devices used in this study. To test the operation reliability, we investigate the gate/SOT pulse configurations and overlays, and we find that an extended gate duration is able to preserve maximized gate benefit and selectivity. Further, the device-scaling criteria are proposed, and we reveal that VGSOT scheme is of great interest as it can mitigate the complex material requirements of achieving high SOT and VCMA parameters for scaled MTJs. Finally, we perform design-to-technology co-optimization analysis to show that VGSOT-MRAM can enable high-density arrays close to two terminal geometries, with high-speed performance and low-power operation, showing great potential for embedded memories as well as in-memory computing applications at advanced technology nodes.

## I. INTRODUCTION

Non-volatile memory (NVM) is believed to address the large stand-by power issues in advanced technology nodes. Among proposed candidates, magnetic random access memory (MRAM) is attracting great attention due to its CMOS process compatibility, high-density, low-power, and relatively fast speed [1]. Recent progress has optimized spin-transfer torque (STT)-MRAM for embedded last-level cache (LLC), micro control unit (MCU), and embedded Flash applications [2-5], with first commercial products beginning to appear in the market. Despite its excellent performance, present STT-MRAM technology is not suited to be implemented in the higher memory hierarchy, such as L1/2 cache memory, due to the significant incubation delay of the STT orientation and the voltage breakdown/writing speed tradeoff. These constraints inhibit reliable programming of perpendicular magnetic tunnel junction (pMTJ) devices at below 5ns [6].

To mitigate the STT limitations, the spin-orbit torque (SOT) [7] and the voltage control of magnetic anisotropy (VCMA) [8] effects have been proposed as alternative writing mechanisms for the next MRAM generations. SOT is the transfer of orbital angular momentum from the lattice to the spin system. When injecting in-plane charge current in a heavy metal layer, the spin Hall effect generates a spin current that would impose a damping-like torque on the neighboring magnetization to induce switching. Simultaneously, a field-like torque induced by Rashba interaction contributes to magnetization reversal acceleration. These led to the concept of SOT-MRAM, a three-terminal device allowing for separated read and write channels which significantly improves the device's endurance. Recent developments have enabled SOT-pMTJ devices with energy-efficient and reliable sub-ns writing capabilities [9,10].

VCMA, on the other hand, promises significant advances towards ultra-low power MRAM. The electronic-based VCMA effect occurs by doping and redistributing electrons at the ferromagnet(FM)/MgO interface [11], which instantly modifies the interfacial magnetic anisotropy of the FM. In a pMTJ, the VCMA-induced free-layer (FL) switching is accomplished by applying a voltage across the MgO tunnel barrier to remove the energy barrier between the parallel (P) and anti-parallel (AP) states, meanwhile an external applied in-plane magnetic field ($B_x$) is required to induce FL precession around its axis [12]. Such switching mechanism is uni-polar and mostly voltage-driven, allowing for ultra-low energy consumption (fJ) at sub-ns writing speeds [13].

However, SOT- and VCMA-pMTJ also present technological challenges. Both require the presence of an in-plane magnetic field to ensure deterministic switching. For SOT, various credible solutions have been proposed and demonstrated, such as integrating an embedded magnet as a field generator [14,15], making use of exchange bias by direct coupling with an antiferromagnet [16,17], or assisting reversal with STT [18]. Therefore, the present challenges for SOT are mostly related to the array density and write efficiency. Indeed, two transistors must be incorporated into a unit cell to operate a 3-terminal SOT device. Meanwhile, the SOT write current remains larger than that for STT-MRAM at the same technology node, which imposes a large selector transistor to accommodate its current, calling for the introduction of new materials with larger spin Hall angles ($\theta_{SH}$) [19,20]. For VCMA-MRAM, the challenges mainly fall on the write margin and efficiency. The VCMA write duration is small, and it is subject to technology variations such as VCMA coefficient ($\xi$), field amplitude, and MTJ diameter distribution, making uniform writing difficult in dense arrays. Furthermore, a large $\xi > 800$fJ/Vm is mandatory to avoid compromising the retention of sub-30nm pMTJ, which remains a major challenge, as typical values in pMTJs are in the range of 30-70 fJ/Vm at device level [21].

To overcome the above-cited issues, a hybrid device combining advantageously SOT and VCMA effects has been proposed [22], namely voltage-gate assisted spin-orbit torque (VGSOT)-MRAM; the cell structure is shown in Fig. 1(a). In the VGSOT concept, SOT is responsible for FL switching while VCMA top-gate assists it. Several advantages emerge by having the VCMA gate assistance. First, it allows for SOT switching at a lower current as it should be reduced proportionally to the decrease in FL-PMA. Second, the VCMA top gate can serve as the MTJ selector, which enables a multi-pillar (MP) cell structure as shown in Fig. 1(b). Such design, with appropriate technology engineering, can also effectively reduce the VGSOT-MRAM cell size to address the density limitation of SOT technologies. Third, by applying a read voltage opposite to write assistance, PMA is increased which further limits the read disturbance [23]. Fourth, the VGSOT design can mitigate the need for achieving challenging high

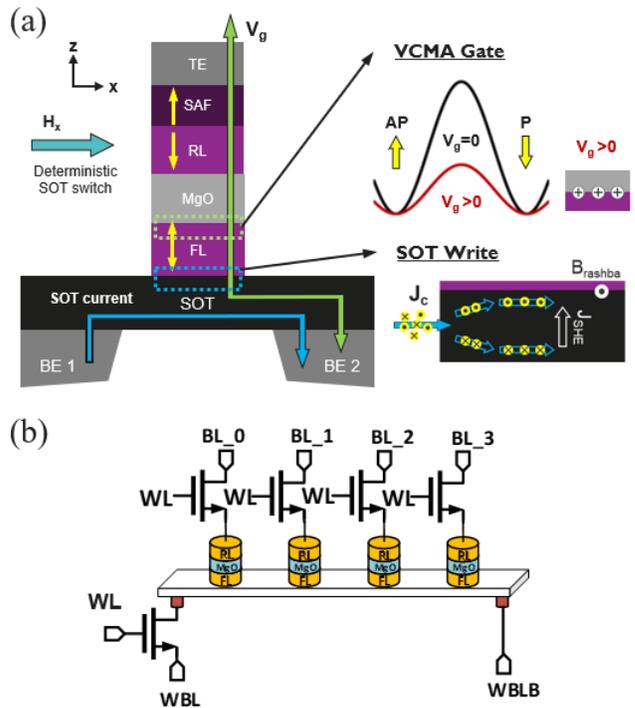

FIG. 1. (a) Illustration of the VGSOT writing scheme. A positive VCMA top gate induces electron accumulation at the FL/MgO interface which lowers the PMA of the FL to allow for SOT switching with a reduced current. (b) Multi-pillar VGSOT cell structure with targeted 4MTJ on a SOT track. Gate voltage serves as the MTJ selector to enable a selective switch.

SOT and VCMA efficiencies while maintaining the advantages of low-power, high speed, and inherent non-volatility.

In this work, we study the VGSOT device properties using CMOS-process compatible pMTJ devices. Section II describes the experimental details, including the materials used for the MTJ stack, device fabrication process, and electrical measurements system. In Section III, VGSOT switching properties are detailed: the VCMA coefficient is quantified by both magnetic field and current methods, and its dependences on size and pulse duration are investigated. Further, the impact of SOT/gate pulse configurations and delays are tested to evaluate the writing scheme for practical operation, and we demonstrate the reliability and resilience of VGSOT concept. Finally, in Section IV, we discuss the design perspectives for scaled VGSOT devices. Based on the proposed MTJ properties requirements, a design-to-technology co-optimization analysis is performed to evaluate VGSOT-MRAM against other embedded memory technologies at 5nm technology node.

## II. EXPERIMENTS

The pMTJ stack is deposited and integrated using our 300mm SOT-MRAM platform [10]. Typical structures consist of a pre-patterned substrate with smoothened bottom electrodes (BE), allowing to electrically contact a 3.5nm tungsten (W) SOT track. The top-pinned pMTJ stack deposited on top of the SOT layer is composed of a 0.9nm CoFeB free-layer (FL), a 1.7nm MgO barrier, and a CoFeB/spacer/Co reference-layer (RL). Through a Ru spacer, RL is anti-ferromagnetically coupled to a [Co/Pt]$_x$ multilayer-based hard-layer (HL), forming the synthetic anti-ferromagnet (SAF) structure. The designed MgO thickness targets a resistance-area (RA) product of 5kΩ·μm$^2$ to suppress the STT current in the gate channel. All layers are sputter-deposited in a Canon Anelva EC7800 cluster tool. After deposition, the stack is annealed at 300°C for 30 minutes in vacuum in a TEL-MSL MRT5000 batch-annealing system, with a 1T out-of-plane magnetic field applied. To pattern the device, we use 193nm immersion lithography, and ion beam etch forms 80-150nm circular MTJ pillars, while optimized etch-stop conditions ensure leaving the SOT layer intact. Next, the SOT layer is etched into 190-300nm wide tracks, depending on the MTJ size. Finally, the Cu top electrode (TE) is fabricated to complete the integration process, forming the 3-terminal devices. The saturation magnetization of the FL ($M_{S,FL}$) measured by the Microsense vibrating sample magnetometer (VSM) before patterning is 900kA/m.

For device characterizations, we use an Hprobe MRAM prober solution. The electrical scheme is shown in Fig. 2(a), both SOT and gate channels are individually connected to a source measure unit (SMU) and a pulse generator. It allows for dc and pulsed measurements with controllable dc voltages ($V_{SOT}^{dc}$/$V_g^{dc}$), pulse voltages ($V_{SOT}$/$V_g$), pulse durations ($t_{p,SOT}$/$t_{p,g}$), and time delays between the two channels. Unless specified, the pulses are synchronized and of the same duration ($t_p$). Besides, the system is provided with individual controls over the out-of-plane and in-plane magnetic fields ($H_z$/$H_x$). To estimate the current distribution in the SOT channel, we use a parallel resistance model shown in Fig. 2(b). The total SOT current ($I_{SOT}$) splits into SOT edge current ($I_{edge}$), SOT active switching current ($I_{sw}$), and FL current ($I_{FL}$). Hence, $I_{eff} = I_{sw} + I_{FL}$ accounts for current shunting in FL and defines the effective current required for switching. The resistivities of the FL and SOT layers are estimated to be 120μΩ·cm and 160μΩ·cm, respectively.

## III. VGSOT-MTJ DEVICE PROPERTIES

### A. Fundamental MTJ Properties

In Fig. 3, the fundamental properties of MTJs are presented as a function of electrical dimension (eCD). Statistics are obtained from 60 devices per size. We measure a median tunneling magneto-resistance (TMR) of 120%, and a coercive field ($H_C$) of 40mT that decreases to 25mT in the smallest devices. Compared to our reference devices [15], we notice that a thicker MgO barrier induces lower $H_C$. We attribute such degradation to the MgO deposition conditions, which we believe can be resolved by stack engineering. Next, we estimate the perpendicular magnetic anisotropy (PMA) field ($H_{k,eff}$) and the thermal stability factor (Δ) using the magnetic field switching probability [$P_{sw}(H_z)$] method [24]:

$$P_{sw}(H_z) = 1 - exp\left\{\frac{-H_{k,eff}f_0\sqrt{\pi}}{2R\sqrt{\Delta}} erfc\left[\sqrt{\Delta}\left(1 - \frac{H_z - H_S}{H_{k,eff}}\right)\right]\right\} \quad (1)$$

where $f_0$ is the attempt frequency, $R$ is the magnetic field sweep rate, and $H_S$ is the effective stray field from the RL/SAF layers. We measure a median $\mu_0 H_{k,eff}$ of 70mT,

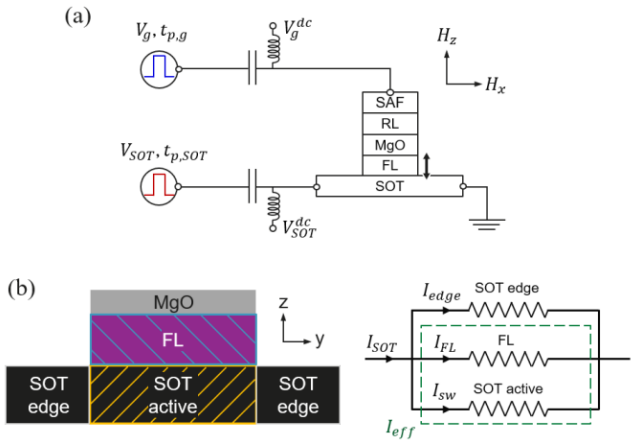

FIG. 2. (a) Schematic of the electrical setup for VGSOT measurements. (b) The parallel resistance model used to estimate the distribution of applied SOT current.

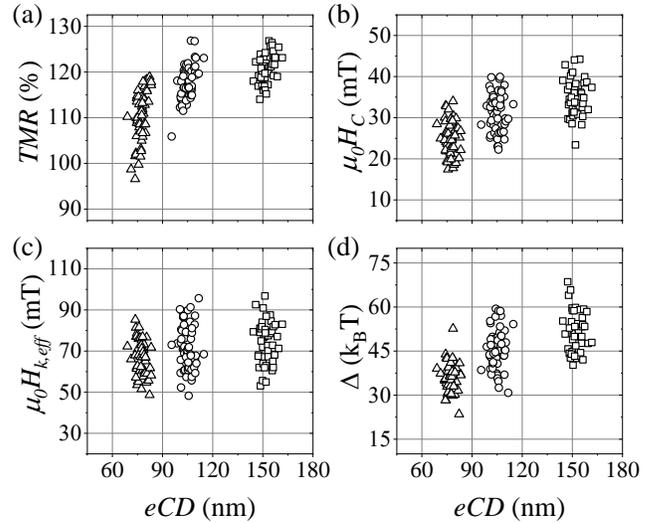

FIG. 3. pMTJ devices properties as a function of eCD for (a) tunneling magneto-resistance (TMR), (b) coercive field ($\mu_0 H_C$), (c) effective perpendicular anisotropy field ($\mu_0 H_{k,eff}$), and (d) thermal stability factor (Δ).

independent of eCD, and a thermal stability factor (Δ) increasing with eCD from 35 to 50 k$_B$T.

To estimate the VCMA coefficient, we first quantify the dependence of $H_{k,eff}$ on dc bias voltage, i.e. the VCMA slope $\frac{\partial H_{k,eff}}{\partial V_g^{dc}}$. The magnetic field switching probability curves are measured under different $V_g^{dc}$ values, as shown in Fig. 4(a). A clear shrinkage (expansion) in the switching fields is observed under positive (negative) $V_g^{dc}$. By fitting the probability curves with Eq. 1, we summarize in Fig. 4(b) the dependences of $\mu_0 H_C$ and $\mu_0 H_{k,eff}$ on $V_g^{dc}$. As expected from the regular VCMA effect, both quantities are reduced (enhanced) for $V_g^{dc} > 0$ ($< 0$) due to electron accumulation (depletion) at the FL/MgO interface. Such VCMA polarity is consistent with the similar stacks in previous studies [25]. Here, we quantify the VCMA slope as 20mT/V. Then, the field estimated VCMA coefficient, denoted as $\xi_H$, is calculated as [26]:

$$\xi_H = \frac{\mu_0 M_{S,FL} t_{FL} t_{MgO}}{2} \frac{\partial H_{k,eff}}{\partial V_g^{dc}} \quad (2)$$

where $t_{FL}$ is the FL thickness and $t_{MgO}$ is the MgO thickness. We obtain median $\xi_H$ = 15fJ/Vm that is independent of eCD [Figure 4(c)]. This lower than typical $\xi_H$ can also be attributed to the MgO deposition conditions that seemly also impacts $H_C$.

## B. Voltage-Gate Assisted SOT Switching

In the following, we discuss the switching results obtained from devices with eCD ~ 80nm sitting on top of 190nm wide SOT track. The SOT channel resistance ($R_{SOT}$) is 320Ω. All measurements are performed at $\mu_0 H_x$ = 10mT, and $\mu_0 H_S$ is compensated by a $\mu_0 H_z$ of -15mT. Figure 5(a) shows exemplary SOT switching probability ($P_{sw}$) curves at $t_p$ = 0.4ns as a function of $V_{SOT}$ under different gate values, obtained from 100 switching events. We observe that VCMA induces a clear decrease (increase) in switching voltage under $V_g$ of 1V (-1V) for both AP-P and P-AP transitions. The critical switching voltage, defined at $P_{sw}$ = 50 %, is converted into critical switching current ($I_c$) and critical current density ($J_c$). Based on Fig. 2(b), a correction factor of 0.49 is applied to $I_c$ for estimating the current contributing to switching $I_{eff}$, i.e. $I_{eff} = 0.49 I_c$, and this will be used in Section IV to project the switching current for optimized processed devices. Figure 5(b) summarizes $I_c$ as a function of $1/t_p$. It shows a typical linear scaling in the sub-ns regime for both transitions at all gate values, implying that the magnetization reversal remains mediated by domain wall nucleation and propagation, and that VCMA gate does not alter the mechanism. Such linear scaling is expressed as [27]:

$$I_c(V_g) = I_{c0}(V_g) + \frac{q(V_g)}{t_p} \quad (3)$$

where $I_{c0}$ is the intrinsic critical current, and $q$ is an effective charge parameter representing the number of electrons needed to be injected into the system before reversing the magnetization, which also indicates the efficiency of angular momentum being transferred from the spin current to the system [27]. We plot the averaged intrinsic critical current ($I_{c0}^{avg}$) and $q$ parameter ($q^{avg}$), averaging from the two switching directions, as a function of $V_g$ in Fig. 5(c) and 5(d), respectively. Both quantities are linearly reduced (increased) for $V_g > 0$ ($< 0$). We obtain $I_{c0}^{avg}(0) = 0.32$mA and $q^{avg}(0) = 1.35*10^{-13}$C, and their dependences on $V_g$ are -49.6μA/V and -5.43*10$^{-14}$C/V, respectively. We interpret the $I_{c0}^{avg}(V_g)$ linear scaling as a reflection of direct modification on the FL-PMA upon gate voltage applications. Since $q$ is understood as describing the conservation of angular momentum [27,28], one can expect that $q^{avg}(V_g)$ should follow $I_{c0}^{avg}(V_g)$ linear trend, as we report in Fig. 5(d). With a 1V gate assistance, we quantify a 25% reduction in $I_{c0}^{avg}$.

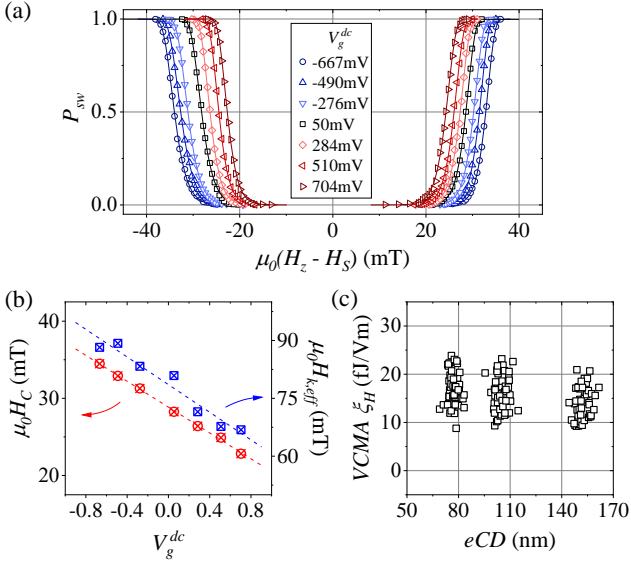

FIG. 4. Representative magnetic field switching probability curves as a function of out-of-plane field for various dc gate voltages ($V_g^{dc}$). (b) $\mu_0 H_C$ and $\mu_0 H_{k,eff}$ as a function of $V_g^{dc}$. $\mu_0 H_{k,eff}$ is obtained by fitting the distributions in (a). (c) VCMA coefficient ($\xi_H$) estimated by field sweep method, showing to be independent of eCD.

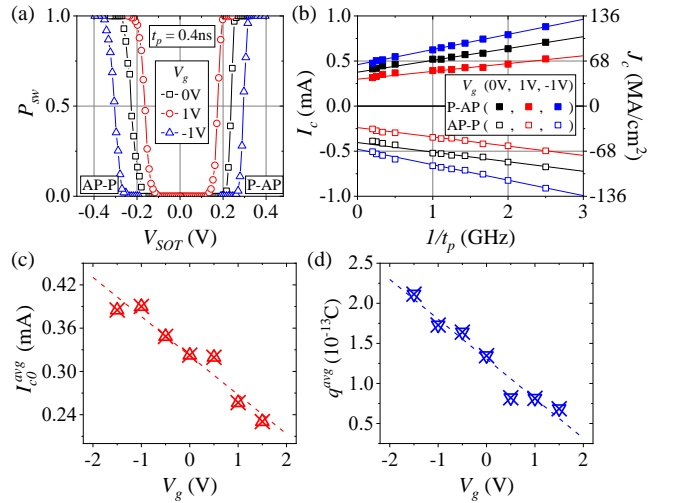

FIG. 5. Classical SOT switching probability distribution ($P_{sw}$) as a function of SOT pulsed voltage ($V_{SOT}$) for different gate values ($V_g$) at $t_p$ = 0.4ns. (b) Critical SOT switching current ($I_c$) at $P_{sw}$ = 50% as a function of $1/t_p$. (c) Averaged intrinsic critical current ($I_{c0}^{avg}$) and (d) averaged $q$ parameter ($q^{avg}$) as a function of $V_g$.

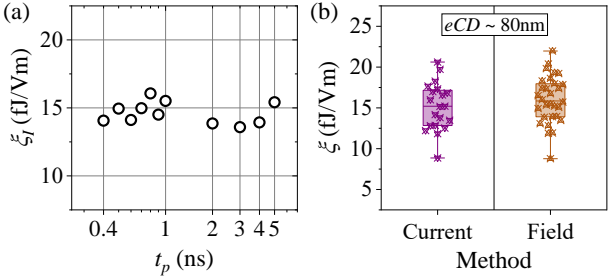

FIG. 6. (a) VCMA coefficient estimated from VGSOT switching current ($\xi_I$) as a function of $t_p$. (e) Comparison of the VCMA coefficients estimated with current switch and field switch methods from 80nm devices.

We also estimate the VCMA coefficient at various $t_p$ based on the VGSOT switching results, denoted as $\xi_I$. The equation to calculate $\xi_I$ is derived from the macro-spin SOT switching current equation [29]:

$$I_c \sim \frac{2e\mu_0 M_{S,FL} t_{FL}}{\hbar \theta_{SH}} \left(\frac{H_{k,eff}}{2} - \frac{H_x}{\sqrt{2}}\right) w_{SOT} t_{SOT} \quad (4)$$

where $\theta_{SH}$ is the spin Hall angle, $w_{SOT}$ is the SOT track width and $t_{SOT}$ is the SOT layer thickness. Since we observe a linear variation of $I_c$ on $V_g$, the slope is expressed as:

$$\frac{\partial I_c}{\partial V_g} = \frac{e\mu_0 M_{S,FL} t_{FL}}{\hbar \theta_{SH}} \frac{\partial H_{k,eff}}{\partial V_g} w_{SOT} t_{SOT} \quad (5)$$

By substituting Eq. 4 and 5 to 2, $\xi_I$ is derived as:

$$\xi_I \sim \frac{\mu_0 M_{S,FL} t_{FL} t_{MgO}}{2} \frac{\partial I_c}{\partial V_g} \frac{H_{k,eff}(V_g=0)}{I_c(V_g=0)} \quad (6)$$

which is an equivalent expression to Ref. [9]. Eq. 6 is typically valid for estimating $\xi_I$ from $I_{c0}$. However, since $\partial I_c/\partial V_g$ slope is normalized by the SOT current and $H_{k,eff}$ at $V_g = 0$V, one can anticipate to find Eq. 6 also applicable to $I_c$, which we confirm in Fig. 6(a), with $\xi_I \sim$ 15fJ/Vm being independent of $t_p$. This further indicates that VCMA in our MTJ is, as expected, an instantaneous and electronic-based effect. Importantly, we find that $\xi_I$ estimation method gives the same median value as the field sweep method $\xi_H$, as shown in Fig. 6(b).

Reduction in switching current and thereby the energy is one of the major benefits of VGSOT switching. To estimate the switching energy, we use an equivalent circuit as shown in Fig. 7(a). We consider here the total energy ($E_{total}$) spent in the present device as the sum of energy dissipations in the SOT channel ($E_{SOT}$) and the gate channel ($E_{gate}$):

$$E_{total} = E_{SOT} + E_{gate} \quad (7)$$

with

$$E_{SOT} = I_c^2 R_{SOT} t_p \quad (8)$$

$$E_{gate} = \frac{V_g^2}{R_{MTJ}+0.5R_{SOT}} t_p \quad (9)$$

where $R_{MTJ}$ is the resistance of the MTJ, taken as the P-state resistance. Figure 7(b) presents the energy values at $t_p = 0.4$ns and 1ns. Since $E_{gate}$ is negligible due to high $R_{MTJ}$, the main energy dissipation comes from $E_{SOT}$. This demonstrates the benefit of the gate assistance to SOT switching, with a $E_{total}$ reduced by 45% at $V_g = 1$V to 30fJ and 41fJ for $t_p = 0.4$ns and 1ns, respectively.

### C. Reliability

In this section, we investigate the reliability of VGSOT switching. First, we study how the gate duration and timing are modifying the averaged critical switching current ($I_c^{avg}$) at $t_{p,SOT} = 5$ns by looking at the gate overlay for $V_g$ = 1V. SOT and gate pulses are synchronized, and $t_{p,g}$ is varied from 0ns to 9ns, as shown in Fig. 8(a). We observe that $I_c^{avg}$ decreases progressively with increasing $t_{p,g}$, and it reaches a maximum reduction of 25% when the SOT pulse is completely overlaid by gate, i.e. $t_{p,g} = t_{p,SOT}$. We interpret this behavior as the decrease of nucleation energy which can also result in a reduction of the nucleation delay time reported in [9,28]. Further, we extend the gate duration prior to the SOT pulse in addition to the primitive 5ns, as shown in Fig. 8(b). It demonstrates that the pre-SOT gate does not influence $I_c^{avg}$. Similar results are also obtained for $t_{p,SOT} = 1$ns (not shown). From the perspective of practical operation, these results suggest that one can moderately increase the gate duration to provide sufficient gate margin to compensate the potential offsets in the SOT current arrival time while maintaining the full advantage of current reduction and bit selectivity.

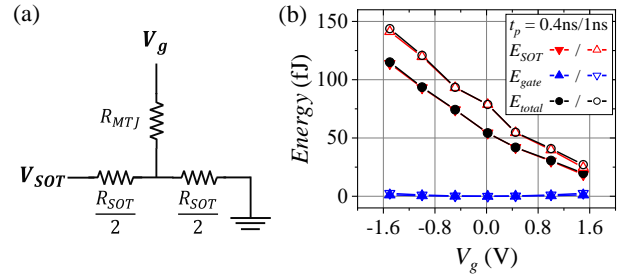

FIG. 7. (a) Schematic of the equivalent circuit used for estimating switching energies. (b) Switching energies as a function of $V_g$ at $t_p = 0.4$ns and = 1ns.

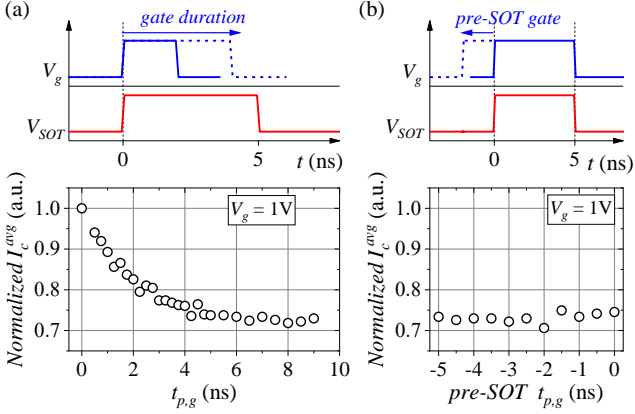

FIG. 8. (a) Normalized $I_c^{avg}$ as a function of $t_{p,g}$ with $V_g = 1V$. SOT and gate pulses are synchronized, and $t_{p,SOT} = 5$ns. $t_{p,g}$ is progressively increased from 0 to 9ns. (b) Normalized $I_c^{avg}$ as a function of additional $t_{p,g}$ prior-to the primitive 5ns overlay of the SOT pulse.

Second, we characterize the write error rate (WER) up to $10^5$ switching events. Figure 9(a) shows the WER of AP-P transition at $t_{p,SOT} = 0.4$ns as a function of $I_{SOT}$ for $V_g =$ - 1V, 0V and 1V. It reveals that even though $V_g = 1V$ reduces $I_c$ by 25%, a wider separation for the WER curves is required to achieve full selectivity. To define gate selectivity, we introduce $S_g$ parameter as being the ratio of WER of the selected cell to read error rate (RER = 1-WER) of the unselected cell:

$$S_g = \frac{WER(V_g)}{RER(V_g=0)} = \frac{WER(V_g)}{1 - WER(V_g=0)} \quad (10)$$

A full gate selectivity requires $S_g > 1$, meaning for exemple that at a WER($V_g$)=1e-5, the RER($V_g$=0) of unselected devices must be below 1e-5. However, we emphasis that the $S_g$ value will mostly depend on application requirements. To predict the required $\xi$ to reach $S_g$ target, we include VCMA effect in the WER switching current distribution model for the SOT scheme [30]:

$$WER(V_g) = \exp\{-f_0 t_p \exp\{-\Delta(V_g)[1 - 2h_s(V_g)(\frac{\pi}{2} - h_s(V_g))]\}\} \quad (11)$$

$$\Delta(V_g) = \Delta(V_g = 0) - \frac{\xi \pi D^2 V_g}{4 k_B T t_{MgO}} = \Delta(V_g = 0) - \beta \xi V_g \quad (12)$$

$$h_s(V_g) = \frac{I_{SOT}}{I_{c0}(V_g)} \quad (13)$$

where $h_s(V_g)$ is the ratio of applied SOT current to the intrinsic switching current, and $\Delta(V_g)$ the retention at applied gate. We note that Eq. 11 typically applies for long $t_p$, and cannot properly quantify actual device parameters. However, it fits qualitatively well our experimental data using $f_0 = 10$GHz, and it can be used to empirically project WER at different $\xi$. We obtain $\Delta(V_g = 0) = 14$ and $\beta = 0.17$m/fJ by fitting the experimental data, and Eq. 11 allows to estimate for that presented devices in this study would require $\xi > 35$fJ/Vm at $V_g = 1V$ to realize $S_g > 1$. This is being largely achievable in standard VCMA-MTJ devices [14,25].

Lastly, we test the endurance for $10^{12}$ cycles at a repetition rate of 50MHz using the following conditions: $I_{SOT} = 1.34$mA and $t_p = 0.4$ns. Such $I_{SOT}$ is much larger than the required current for WER < $10^{-5}$. Figure 9(b) and 9(c) show that both the selected ($V_g = 1V$) and unselected ($V_g = 0V$) cells sustain this intensive writing stress, and that both MTJs and SOT tracks remain intact without any degradation in resistances, proving that VGSOT writing scheme is robust and resilient for high performance memory applications.

## IV. VGSOT DEVICE SCALING PERSPECTIVES

### A. Device scaling criteria

From above studies, we project the required $\theta_{SH}$ and $\xi$ for the scaled VGSOT-pMTJ based on two scaling criteria: SOT critical current and gate selectivity. We consider a 30nm MTJ having $\Delta = 60$ and RA = $20\Omega \cdot \mu m^2$ ($t_{MgO}$~1nm); these conditions are defined to achieve the required retention for embedded memory applications and to minimize the reading latency. Notice that the STT effect at such RA range would have negligible impact on the magnetization when writing at sub-ns speeds [31,32]. By combining Eq. 2 and 4, $I_c$ under gate assistance is derived as:

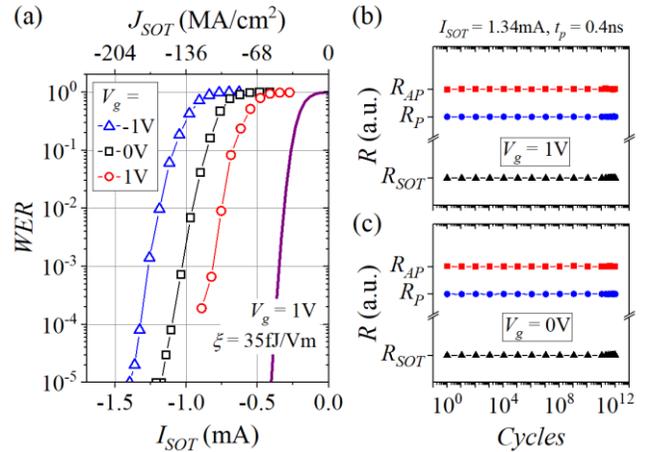

FIG. 9. (a) WER of $10^5$ events for the AP-P transition at $t_p = 0.4$ns. Purple curve predicts the WER for $\xi = 35$fJ/Vm at $V_g = 1V$, which is required to achieve full selectivity. (b) Endurance tests of $10^{12}$ cycles at applied SOT current $I_{SOT} = 1.34$mA and $t_p = 0.4$ns with 50MHz repetition rate for (b) selected cell $V_g = 1V$ and (c) unselected cell $V_g = 0V$.

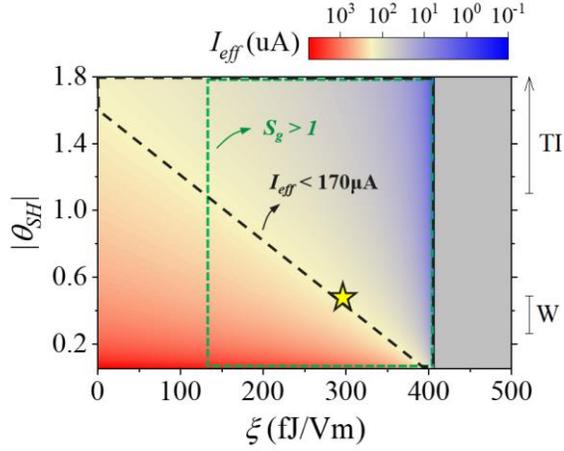

FIG. 10. Effective SOT switching current as a function of $\theta_{SH}$ and $\xi$. The following parameters are applied: $D$ = 30nm, $\Delta$ = 60, $t_{MgO}$ = 1nm, and $V_g$ = -0.7V. Black outlined area indicates the $I_{eff}$ < 170μA and green outlined area indicates $S_g$ > 1. Star symbol exemplifies the required $\theta_{SH}$ and $\xi$ to fulfill these targets.

$$I_c(V_g) \sim \frac{2e\mu_0 M_{S,FL} t_{FL}}{\hbar \theta_{SH}} \left( \frac{H_{k,eff}}{2} - \frac{\xi V_g}{\mu_0 M_{S,FL} t_{FL} t_{MgO}} - \frac{H_x}{\sqrt{2}} \right) w_{SOT} t_{SOT} \quad (14)$$

Then, we implement $\Delta$ into Eq. 14 and obtain:

$$\Delta = \frac{\mu_0 M_{S,FL} H_{k,eff} \pi D^2 t_{FL}}{8 k_B T} \quad (15)$$

$$I_c(V_g) \sim \frac{w_{SOT} t_{SOT}}{\hbar \theta_{SH}} \left( \frac{8 e k_B T \Delta}{\pi D^2} - \frac{2 e \xi V_g}{t_{MgO}} - \frac{e \mu_0 M_{S,FL} t_{FL} H_x}{\sqrt{2}} \right) \quad (16)$$

where $D$ is the MTJ diameter, $k_B$ is the Boltzmann constant, and $T$ is the ambient temperature. Using Eq. 16 and the parallel resistance model, we project in Fig. 10 the required $I_{eff}$ for 1ns writing speed based on our experimental results, with $V_g$ being fixed at -0.7V to adapt the maximum supply voltage of the core transistor at beyond 10nm technology node [33]. Moreover, to achieve a fully-functional VGSOT-MRAM, we consider $S_g$ > 1 at WER($S_g$) = 10$^{-5}$, which corresponds to $\xi$ > 130fJ/Vm in the scaled MTJ.

With these calculations, we find that VGSOT scheme can mitigate the requirements of challenging SOT and VCMA parameters. For example, one can achieve $I_{eff}$ = 170μA using the standard material in SOT-MRAM technology such as tungsten with $\theta_{SH}$ = -0.45 [34]. Correspondingly, a $\xi$ = 300fJ/Vm is needed, which can be implemented by appropriate interface engineering [35]. As a comparison, these values are much more relaxed than the requirements for SOT-MRAM ($|\theta_{SH}|$ > 1.6) and for VCMA-MRAM ($\xi$ > 800fJ/Vm, considering $t_{MgO}$ = 1.5nm at $\mu_0 H_x$ = 0mT) of the same $\Delta$. Further reduction in $I_{eff}$ would require emerging SOT materials with larger values such as topological insulators (TI) [19,20]. We remark that $\xi$ > 400fJ/Vm is not desired as it would enter into a pure VCMA operation regime.

### B. Design-to-Technology Co-Optimization Analysis

Finally, we perform an extensive design-to-technology co-optimization (DTCO) analysis to benchmark the VGSOT performance at 5nm technology node against other embedded memory technologies such as static-RAM (SRAM), STT-MRAM, and SOT-MRAM. In Fig. 11, the performance values for SRAM high-density (HD) and high-performance (HP) are estimated based on the device sizing constraints of different foundries [36,37,38,39]. We consider a 45nm gate pitch and a 30nm metal pitch for our STT-MRAM [40] and SOT-MRAM [41] designs to optimize the performances. The power consumptions are estimated for 128kbit memory macro with 64bit data output. For VGSOT-MRAM, the 2 and 4 MTJ variants in VGSOT represent the number of pillars on a single SOT track within a single bit-cell [Fig. 1(b)]. VGSOT performance estimation would include variables such as number of MTJ pillars, SOT line length/resistance, spin Hall angle, and VCMA coefficient. We estimate that the VGSOT-4MTJ design can significantly reduce the effective area compared to the conventional SOT-MRAM, making it comparable to STT-MRAM and less than 50% of the HD-SRAM at minimal performance degradation. As emphasized earlier, due to the large MTJ resistance in our present devices, the read performance is limited. This can be resolved by optimizing the RA product to RA = 20Ω·μm$^2$ without compromising the benefits of VCMA gate effects. The write power of 2MTJ and 4MTJ designs under the present VCMA coefficient remains larger than the typical SOT cell, which is due to the increase in SOT channel length and resistance. The minor difference between the write power of 2MTJ and 4MTJ designs stems from the change in the word-line (WL) capacitance when increasing the number of pillars. Eventually, the write performance can also be enhanced with improved SOT and VCMA parameters. We project the VGSOT performance based on the following conditions: $\Delta = 60$, RA = $20\Omega\cdot\mu m^2$, SOT resistivity $\rho_{SOT} = 160\mu\Omega\cdot cm$, $\theta_{SH} = -0.45$, and $\xi = 300fJ/Vm$. It shows 2x and 10x improvements, respectively, in read latency and write power, opening design perspectives for high-performance, low-power, and high-density embedded memories as well as in-memory computing applications.

| Specs | SRAM | | STT | SOT | VGSOT | | |
|---|---|---|---|---|---|---|---|
| | HD | HP | | | 2MTJ | 4MTJ | Projected |
| Area ($\mu m^2$) | 0.021 | 0.028 | 0.0084 | 0.016 | 0.0122 | 0.009 | 0.009 |
| RD Power/bit (nW) | 7.28 | 18.7 | 6.20 | 25.5 | 3.16 | 2.96 | 5.8 |
| WR Power/bit (nW) | 9.85 | 25.7 | 24.4 | 31.4 | 47.5 | 46.5 | 4.6 |
| RD Latency (ns) | ~1.5 | ~0.8 | ~2.8 | ~1 | ~10 | ~10 | ~5 |
| WR Latency (ns) | ~1.5 | ~0.8 | ~20 | ~2 | ~1 | ~1 | ~1 |
| $V_{DD}$ (V) | ~0.7 | ~0.7 | ~0.7 | ~0.7 | ~0.7 | ~0.7 | ~0.7 |
| Retention | - | - | >10yrs | >10yrs | >10yrs | >10yrs | >10yrs |
| Endurance | >10$^{16}$ | >10$^{16}$ | >10$^9$ | >10$^{14}$ | >10$^{14}$ | >10$^{14}$ | >10$^{14}$ |

FIG. 11. DTCO analysis of performance for different embedded memory technologies at 5nm technology node, including SRAM, STT-MRAM, SOT-MRAM and VGSOT-MRAM. Projected values for VGSOT are estimated with improved MTJ properties.

## V. CONCLUSION

We presented a complete voltage-gate assisted spin-orbit torque switching study on pMTJ devices. Such writing approach is of great interest on one hand to minimize the SOT write energy, and on the other hand to gain architecture density by combining multiple MTJs on the same SOT track with VCMA gate acting as the bit selector. From the dependences of SOT critical switching current on gate voltage (with a gate duration equal to SOT pulse), we found that both intrinsic current and charge conservation parameter $q$ are scaling linearly with the gate voltage, which corresponds to the expected direct modification of PMA and nucleation energy by VCMA. We also propose a generalized method to evaluate the VCMA coefficient from time-dependent switching experiments. As expected from pure electronic VCMA effect, $\xi_I$ is independent of writing speed, and its value is similar to that obtained from the traditional magnetic field method $\xi_H$. In addition, we showed that a fully gate-overlaid SOT pulse is required to maximize the gate effects, and both pre-SOT and post-SOT gate have no impact on the SOT switching efficiency or selectivity. To highlight the VGSOT switching reliability, we investigated the write error rate down to $10^{-5}$ with no error, and both selected and unselected cell endured intensive write stress for more than $10^{12}$ cycles without signature of degradation. Interestingly, we showed through simple device metric analysis that VGSOT scheme can ease the requirements of challenging high SOT and VCMA parameters. This would relieve the need for new complex SOT materials with $\theta_{SH} > 1$, which currently raises the complexities for optimizing MTJ properties and integration process. Finally, we benchmarked the VGSOT device performance against other embedded memories (SRAM, STT-MRAM, SOT-MRAM) using design-to-technology co-optimization analysis, and we projected VGSOT-4MTJ design to be energy efficient with fast read/write speed and density close to a 2-terminal device. This makes VGSOT an appealing candidate for memory applications requiring high-density, high performance, non-volatility, and low-power.

## ACKNOWLEDGMENT

This work was supported by IMEC's Industrial Affiliation Program on MRAM device.